\documentclass[conference,letterpaper]{IEEEtran}

\InputIfFileExists{synctex.tex}{%
\synctex=1
}{%
}

\usepackage{xcolor}
\usepackage{cite}
\usepackage{amsmath}
\usepackage{amssymb}
\usepackage{url}
\usepackage[linesnumbered,ruled,noend]{algorithm2e}
\usepackage[pdftex]{graphicx}
\usepackage{wrapfig}
\usepackage{tikz}
\usepackage{balance}
\usepackage{paralist}
\usepackage[caption=false]{subfig}
\usepackage{array}
\usetikzlibrary{arrows,automata}

\newcommand{\ol}[1]{\textcolor{blue}{\ifmmode \text{[OL: #1]}\else [OL: #1] \fi}}
\newcommand{\vh}[1]{\textcolor{orange}{\ifmmode \text{[VH: #1]}\else [VH: #1] \fi}}
\newcommand{\tv}[1]{\textcolor{magenta}{\ifmmode \text{[TV: #1]}\else [TV: #1] \fi}}
\newcommand{\lh}[1]{\textcolor{green}{\ifmmode \text{[LH: #1]}\else [LH: #1] \fi}}
\newcommand{\jm}[1]{\textcolor{red}{\ifmmode \text{[JM: #1]}\else [JM: #1] \fi}}

\newcommand{\secref}[1]{Section~{\ref{#1}}}

\newcommand{\calS}{\mathcal{S}}

\newcommand{\A}[0]{\mathcal{A}}
\newcommand{\bbA}[0]{\mathbb{A}}
\newcommand{\bbN}[0]{\mathbb{N}}

\newcommand{\lang}{L}

\newcommand{\langA}[0]{L_\A}

\newcommand{\trans}[0]{\delta}
\newcommand{\extrans}[0]{\hat{\trans}}

\newcommand{\errorAc}[0]{\mathit{error}_{ac}}

\newcommand{\ap}{\mathit{AP}}

\newcommand{\atp}{A_{\mathit{TP}}}
\newcommand{\afp}{A_{\mathit{FP}}}

\newcommand{\prob}{\mathit{Prob}}

\newcommand{\lab}[0]{\ell}

\newcommand{\distance}[0]{\mathit{dist}}
\newcommand{\mdistance}[0]{D}
\newcommand{\disteq}[0]{\sim_\mdistance}
\newcommand{\frequency}[0]{\mathit{freq}}
\newcommand{\mfrequency}[0]{F}
\newcommand{\prunedA}{\A_R}
\newcommand{\reduce}[0]{\textsc{Reduce}\xspace}

\newcommand{\snort}[0]{\textsc{Snort}\xspace}

\newcommand{\netbench}[0]{\textsc{Netbench}\xspace}
\newcommand{\suricata}[0]{\textsc{Suricata}\xspace}
\newcommand{\bro}[0]{\textsc{Bro}\xspace}

\newcommand{\bigO}[0]{\mathcal{O}}
\newcommand{\bigOof}[1]{\bigO(#1)}

\newcommand{\lut}[0]{\mathit{LUT}}
\newcommand{\lutof}[1]{\lut(#1)}
\newcommand{\accpt}[0]{\mathit{Acc}}
\newcommand{\accptof}[1]{\accpt(#1)}
\newcommand{\rangeof}[1]{[#1]}
\newcommand{\varof}[1]{\langle \text{#1} \rangle}

\newcommand{\qed}[0]{\hfill$\Box$}

\newcommand{\red}[1]{\textcolor{red}{#1}}

\begin{document}

\title{Deep Packet Inspection in FPGAs via\\ Approximate Nondeterministic Automata }


\author{
\IEEEauthorblockN{
  Milan \v{C}e\v{s}ka,
  Vojt\v{e}ch Havlena,
  Luk\'{a}\v{s} Hol\'{i}k,
  Jan Ko\v{r}enek,\\
  Ond\v{r}ej Leng\'{a}l,
  Denis Matou\v{s}ek,
  Ji\v{r}\'{i} Matou\v{s}ek,
  Jakub Semri\v{c}, and
  Tom\'{a}\v{s} Vojnar
}
\vspace{2mm}
\IEEEauthorblockA{
  Brno University of Technology,
  Faculty of Information Technology,
  IT4I Centre of Excellence,
  Czech Republic
}}

\maketitle

\begin{abstract}

  Deep packet inspection via regular expression (RE) matching is a~crucial task of
  network intrusion detection systems (IDSes), which secure Internet
connection against attacks and suspicious network traffic.
  Monitoring high-speed computer networks (100\,Gbps and faster) in a~single-box
  solution demands
  that the RE matching, traditionally based on
  \emph{finite automata} (FAs), is accelerated in hardware.
  In this paper, we describe a~novel FPGA architecture for RE matching that is able
  to process network traffic beyond 100\,Gbps.
  The key idea is to reduce the required FPGA resources by leveraging approximate nondeterministic FAs (NFAs).
  The NFAs are compiled
  into a~multi-stage architecture starting with the least precise stage with a~high throughput and
  ending with the most precise stage with a~low throughput.
  To obtain the reduced NFAs, we propose new approximate reduction techniques
  that take into account the profile of the network traffic.
  Our experiments showed that using our
  approach, we were able to perform matching of large sets of REs from \snort, a popular~IDS, on unprecedented network speeds.

\end{abstract}

%


\vspace{-0.0mm}
\section{Introduction}\label{sec:intro}
\vspace{-0.0mm}

Intrusion Detection Systems (IDSes), such as \snort~\cite{Snort},
\suricata~\cite{Suricata}, or \bro~\cite{bro}, are widely used to secure
Internet connection against attacks and malicious traffic.
One of the prominent approaches for IDSes is \emph{deep packet inspection} (DPI), which is based on matching \emph{regular
expressions} (REs) describing attack patterns
against network traffic.
Despite the recent rapid increase of encrypted traffic on the
Internet, RE-based DPI is still in high demand since IDSes are often deployed
at network entry points after decryption.


Due to the increasing number of security vulnerabilities and
network attacks, the number of REs in IDSes is constantly growing.
At the same time, the speed of networks is growing too---telecommunication
companies started to deploy 100\,Gbps links, the 400\,Gbps Ethernet standard has
recently been ratified~\cite{400gbethernet}, and large data centers already call
for a~1\,Tbps technology.
Consequently, despite many proposed optimisations, existing IDSes are
still far from being able to process the traffic in current high-speed networks at the line speed.
The best software-based solution we are aware of is the one in~\cite{VallentinSLLPT07}, which
can achieve a~100\,Gbps
throughput using \bro on a~cluster of
servers with a~well-designed distribution of network traffic.
Processing network traffic at such speeds in single-box IDSes is far beyond
the capabilities of software-based solutions---hardware acceleration is needed.


A~well-suited technology for accelerating IDSes is that of  \emph{field-programmable
gate arrays} (FPGAs).
They provide high computing power and flexibility for network traffic
processing, and they are increasingly being used in data
centers~\cite{configurable-cloud-acceleration,
reconfigurable-datacenter-services} for this purpose.
The flexibility of FPGAs allows them to match REs at speeds over
100\,Gbps~\cite{MatousekKP16}.
Such high speeds, however, put excessive demands on the resources of FPGAs.
The sets of the matched REs are complex, large, and still growing,
and matching on the speeds of tens and hundreds of Gbps
requires massive parallelization.
For instance, in the HW architecture that we propose in \secref{sec:arch},
processing 100\,Gbps input network traffic requires 64
concurrently functioning RE matching units (of 8-bit input width operating at
200\,MHz) and processing 400\,Gbps requires even 256 units.
These demands easily exceed the size of any available FPGA chip.
Reducing the consumed resources is thus of paramount
importance.




The FPGA matching units traditionally implement \emph{finite automata} (FAs),
either \emph{deterministic} (DFAs) or \emph{nondeterministic} (NFAs).
In this paper, we focus on NFAs since they are often much smaller than the
corresponding DFAs and can be efficiently mapped into FPGAs as shown, e.g.,
by~\cite{sidhu-prasanna-nkda,ClarkS03,SourdisBispo2008,LinVLSIprefix,AtMostTwoHot}.

%
%

The main conceptual difference from other works on efficient synthesis of NFAs into FPGAs (described in more detail in the related work section),
which allows us to achieve much higher matching speeds, lies in leveraging \emph{reduced} NFAs that \emph{over-approximate} the language of the original NFA---we design
novel reduction techniques that provide high-quality tradeoffs between the precision and reduction factors.

Subsequently, in order to utilise the resources of an FPGA in the best possible way,  we
propose a~\emph{multi-stage} architecture of the RE matching engine.
Consider an~NFA~$\A$ that recognizes the language~$\lang$ of a~given set of REs.
The proposed architecture is composed of several stages where the first stage
contains many, concurrently running, copies of small NFAs that crudely
over-approximate~$\lang$, and further stages contain a~smaller number of copies
of NFAs that are larger but more precise.
The throughput of the stages decreases:
the first stage needs to be able to process the network traffic at the line
speed while the subsequent stages can handle less.
The task of every stage is to decrease the amount of traffic
entering the next stage by removing some of the packets that are guaranteed not
to be in~$\lang$.
The last stage contains either~$\A$, or, in the case of insufficient FPGA
resources, a reduced version of $\A$ that over-approximates $\lang$ (in which case the
final processing step takes place in software).
The NFAs used in the over-approximating stages need to discard a~significant
portion of their input from further processing while keeping all of the
suspicious traffic.
In the worst case, i.e., when every input packet is a~member of~$\lang$, this
is, naturally, impossible.
Fortunately, in standard traffic on a backbone network,
only a~small portion of the packets is normally in $L$ (the extreme case when
a~significant proportion of network traffic is in~$L$ usually corresponds to
a DoS attack, which needs to be handled by other means than~IDSes).

Our approximate NFA reduction takes an advantage of particularities of standard network traffic.
Namely, given an NFA constructed from the REs of interest,
we label its states with their \emph{significance}---the likelihood that they will be used during processing a packet---,
and then simplify the least significant parts of the automaton.
The simplification is implemented by \emph{pruning} and \emph{merging} of the insignificant states.
The significance of a~state is determined using \emph{training traffic},
a~finite sample of ``standard'' traffic from the network node where the IDS is
to be deployed (it may be necessary to generate a~new design once in a~while).
The reduction scales well---the worst-case time complexity of the most expensive step, computing the state labelling,
is~$\bigOof{n^2k}$ where $n$~is the number of states of the NFA and $k$~is the
size of the training traffic (since the automata are usually sparse,
the quadratic factor is rarely an issue on real-world examples).

We  implemented the proposed approach and evaluated it on
REs taken from the IDS \snort and other resources.
We were able to obtain a~substantial reduction of the size of the NFAs while keeping
the number of false positives low.
When used within the multi-stage architecture, we were able to perform RE
matching at 100 and 400\,Gbps on sets of REs whose sizes were far beyond the
capabilities of existing~solutions.

 The contributions of this paper are the following:
\begin{enumerate}
  \item  We propose a new HW architecture for matching REs in high-speed computer
    networks using several stages of NFAs that over-approximate the matched REs.
  \item  We developed new scalable over-approximation techniques that take into account the profile of network traffic and thus provide great reductions of the NFAs while keeping an acceptable error.
  \item We present experimental results proving that our approach can indeed be
  successfully used to implement single-box hardware-accelerated IDSes
  at speeds of 100\,Gbps or even 400\,Gbps, which is,
  taking into account the size and complexity of the considered REs,
  far beyond the capabilities of current solutions.
\end{enumerate}

\vspace{-0.0mm}
\subsection*{Related Work}\label{sec:related}
\vspace{-0.0mm}

%
Many different architectures for resource-efficient mapping of NFAs for fast RE
matching into FPGAs have been designed, starting with the work of
Prasana and Sidhu~\cite{sidhu-prasanna-nkda}.
Later, Clark and Schimmel~\cite{ClarkS03} reduced hardware
resources by using a~shared decoder of input symbols, and the architecture of Sourdis
\emph{et al.}~\cite{SourdisBispo2008} implements an NFA transition table by
a~pre-decoded CAM.
Lin \emph{et al.}~\cite{LinVLSIprefix} introduced an architecture that allows
 hardware resources to be shared for matching REs with the same prefixes, infixes,
and suffixes.
Furthermore, the work of Yun and Lee~\cite{AtMostTwoHot} improved the encoding
of states using the so-called at-most-two-hot encoding.


To increase the RE matching speed, some architectures make the NFAs process multiple
bytes of the input per clock cycle.
Prasanna \emph{et al.}~\cite{PrasannaREM12} introduced spatial stacking for
multi-character matching, but the high fan-out of
the automaton significantly decreases its frequency, already for matching 8
bytes
per clock cycle.
%
Achieving the 100\,Gbps throughput, which requires more than 64 bytes to be
processed at once, is thus not possible.
%
Becchi and Crowley~\cite{BecchiMultiStriding} introduced a~multi-striding
technique, which is widely used to increase the throughput of
many NFA-based RE matching architectures.
Multi-striding alone, however, cannot increase the processing speed
to 100\,Gbps because with the length of the stride,
the NFA grows
rapidly,
and the frequency drops
dramatically~\cite{MatousekKP16}.

We build on our previous work~\cite{MatousekKP16} introducing parallel pipelined
automata, which can scale the throughput of NFA-based RE matching to over 100\,Gbps.
The processing speed is increased at the cost of
a~linear growth of hardware resources because the architecture consists of many
automata connected in one processing pipeline. In a~recent work, a novel FPGA architecture that significantly improved
the throughput of DFA-based RE matching has been proposed by Yang \emph{et al.}~\cite{FPGAnew}.
The architecture achieves a~throughput of 140\,Gbps on a~single FPGA chip for
IDS modules where the underlying \emph{DFAs} have up to 10k states
(corresponding to 34\,REs of \snort).
Our new multi-stage architecture, which leverages approximate \emph{NFAs},
fundamentally improves the results of~\cite{MatousekKP16} and~\cite{FPGAnew}.
We can achieve, on a~similar chip, a~throughput beyond 200\,Gbps for much more complicated sets of REs
(e.g., \snort's \texttt{spyware} module with 461\,REs where the under\-lying
NFA has $\sim$10k states and the corresponding DFA is prohibitively
large---our attempt of its determinisation depleted\linebreak the available memory (32\,GiB) after reaching 616k states).


The works closest to our NFA reduction techniques are~\cite{LuchaupDJB14}
and~\cite{CeskaHHLV18}.
In~\cite{LuchaupDJB14}, the authors address the issue of software-based acceleration
of matching REs describing network attacks in \snort.
To reduce the number $k$ of membership tests needed for matching a packet against $k$ distinct DFAs,
\cite{LuchaupDJB14} builds a ``search tree'' with the $k$ DFAs in its leaves
and with the inner nodes occupied by preferably small DFAs that over-approximate
the union of their children
(the precise DFA accepting the union is prohibitively large).
Matching a~packet then means to propagate it down the tree as long as it belongs
to languages of the DFA nodes.
The over-approximating DFAs are constructed using a similar notion of
significance of states as in our approach.
The differences from our work are the following:
(1)~\cite{LuchaupDJB14} needs several membership tests per packet,
(2)~it targets only software and its hardware implementation would be too complex,
(3)~it does not consider reduction of NFAs, and
(4)~it uses only a pruning-based reduction (we also employ merging and
simulation-based reductions).

Our previous work in~\cite{CeskaHHLV18} also targets approximate reductions of NFAs used in
HW-accelerated IDSes.
In contrast to the approach in this paper, which uses a~sample of the network traffic to
determine the states to remove, \cite{CeskaHHLV18} uses
a~probabilistic model of the traffic in the form of a~probabilistic automaton,
which is used to label states of the  NFA by probabilities of activation.
The method has the following drawbacks:
(1)~the model can be constructed from a sample of traffic  only semi-automatically, with an
aid of a network expert, and
%
%
(2)~it scales to NFAs of up to only around 1,300 states,
whereas our NFAs are sometimes an order of magnitude larger. Moreover,
\cite{CeskaHHLV18} considers only a~single-stage architecture.

Simulation-based reduction  is a~standard language-preserving NFA reduction technique,
which, apart from its basic form \cite{BustanG00} (used in
\cite{KosarZK13} in the context of FPGA-accelerated IDSes), comes in
a~number of advanced variants (e.g. look-ahead, multi-pebble,
or mediated \cite{ClementeM13,Etessami02,AbdullaCHV14}), many of them
implemented in the tool \reduce~\cite{Reduce}.
Although \reduce is a~part of our workflow,
the power of simulation reduction alone is by far insufficient for our purposes.
%

Finally, we compare our approach with RE-matching techniques that use modern
general-purpose GPUs.
As in FPGAs, we can distinguish architectures leveraging both DFAs and NFAs.
Prominent GPU architectures based on DFAs include Gregex~\cite{Gregex},
hierarchical parallel machines~\cite{GPUNIDS}, and a recent work employing
algorithm/implementation co-optimization based on a~GPU performance
model~\cite{GPUnew}.
These architectures are able to perform RE matching at the theoretical
throughput of 100--150\,Gbps.
Their practical performance is, however, limited by the packet
transfer throughput (e.g., the throughput of \cite{Gregex}\linebreak was 25\,Gbps on
NVIDIA GTX260), and, indeed, the complexity of RE modules they can
handle due to determinization.

An~RE matching architecture for GPUs based on NFAs was proposed in
iNFAnt~\cite{iNFAant} and further improved in~\cite{iNFAantnew}.
In contrast to the aforementioned DFA-based solutions, this architecture can
handle complex RE modules where the underlying NFAs have over 10 thousand
states.
The performance of this work is, however, significantly inferior to our
FPGA-based architecture.
For example, for a category of \snort modules where the underlying NFAs have
\mbox{3--18} thousand states, \cite{iNFAantnew}~reports an overall throughput of
\mbox{1--2\,Gbps} on NVIDIA Tesla c2050.
Our experiments in~\secref{sec:eval} show that, on \snort modules with similar sizes,
our approach is able to achieve a throughput of over 100\,Gbps.

  Further disadvantages of using GPUs are that
  \begin{inparaenum}[(i)]
    \item  they are notorious energy hogs, e.g., the NVIDIA Tesla K20c (used
      by~\cite{GPUnew}) card's TDP is 225\,W, which is significantly higher than
      the 75\,W consumed by our FPGA-based architecture and
    \item  they impose a~high latency on packet processing (in the order of
      hundreds of ms compared to $\sim$10\,$\mu$s of our solution).
  \end{inparaenum}

\vspace{-0.0mm}
\section{Architecture}\label{sec:arch}
\vspace{-0.0mm}

In this section, we first describe the basic architecture of the RE matching
engine from~\cite{MatousekKP16}, which is used as a~building block of our
multi-stage architecture described later.

\vspace{-0.0mm}
\subsection{Regular Expression Matching Engine}\label{sec:reg_engine}
\vspace{-0.0mm}

\begin{figure}[!h]
    \subfloat[\label{fig:pipelined_automata}]
    {
      \hspace*{-2mm}
      \includegraphics[width=46.5mm]{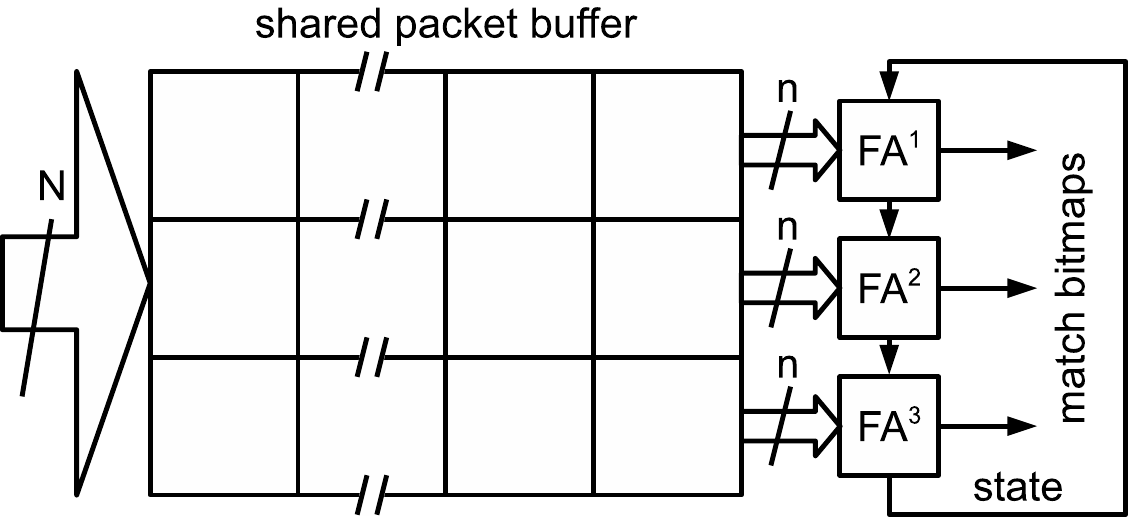}
    }
    ~
    \subfloat[\label{fig:multistage}]
    {
      \includegraphics[width=38mm]{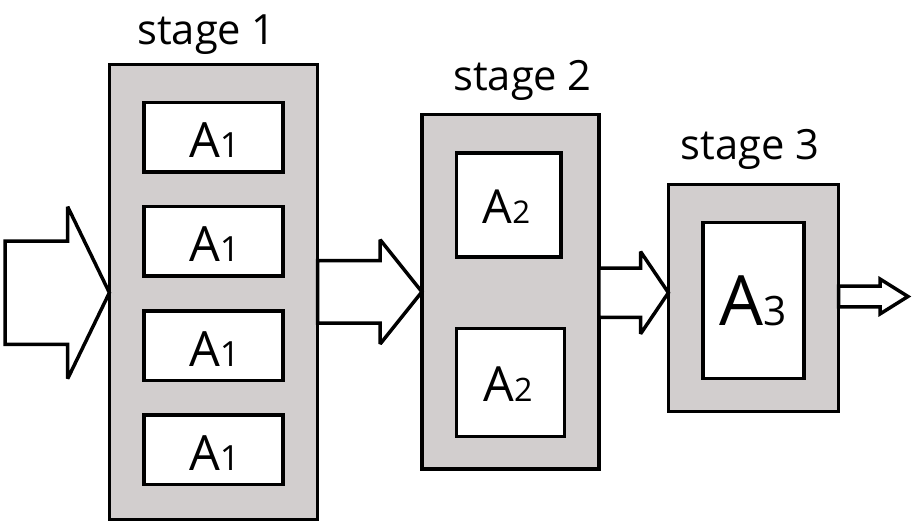}
    }
    \caption{(a) The architecture of an RE matching engine with $k = 3$ pipelined FAs sharing a single packet
             buffer able to store $N$-bit data words (columns), each of which
             consists of $k$~independent $n$-bit blocks (rows). (b) An example of
             the multi-stage architecture with 3 stages.}
\end{figure}

The architecture proposed in~\cite{MatousekKP16} uses \emph{pipelining} to
enable RE matching in high-speed networks on a~single FPGA.
An example of~an instance of this architecture consisting of $k = 3$ pipelined
FAs and a~single
packet buffer is shown in Fig.~\ref{fig:pipelined_automata}.

The packet buffer stores every $N$-bits-wide input data word as $k$ independent data
blocks of $n$ bits (i.e., $N = k \cdot n$).
Each of the $k$~FAs can read data blocks on the corresponding row of the buffer
and also receives the configuration of the previous FA in the pipeline that
processed the previous block of the packet (or the initial configuration if the
FA is processing the first block of a~packet).
Based on these inputs, the FA either computes the next configuration or
determines an RE that matches the packet.
These pieces of information are sent to the next FA in the pipeline until the last block
of the packet is processed, in which case a bitmap encoding all matching REs is
output.
As only one of the $k$~pipelined FAs can perform RE matching in a given packet
at any moment, the whole pipeline can theoretically perform RE matching in
$k$~packets in parallel.

%
The FA that encodes the complete set of REs is replicated $k$~times.
For instance, to achieve the throughput of 100\,Gbps, we
used the replication factor $k=16$ in~\cite{MatousekKP16}.
Such a~massive replication costs a~lot of FPGA resources,
which becomes the major bottleneck of the approach.
On the other hand,
the replication is also a unique opportunity for optimizations
because every reduction of the FA's size means a~16-times higher reduction in
the whole architecture.
%
From this point of view, using NFAs is preferable to DFAs because, especially in the combination with
language-preserving simulation-based reduction, they are much more succinct than
DFAs, and can also be mapped into FPGAs more efficiently than DFAs.
They are, however, still prohibitively large in many practical cases; this calls
for better reduction methods.
%
In~\secref{sec:approx_red}, we present two such methods in the form
of approximate reductions, which do not preserve the language of the FA.
The reduced FAs are used within our multi-stage architecture (built on
top of the pipelined approach of~\cite{MatousekKP16}), which allows us to
flexibly adjust the tradeoff between precision and consumed resources.
The multi-stage architecture is described next.
%
%
%

\vspace{-0.0mm}
\subsection{Multi-Stage Regular Expression Matching Architecture}\label{sec:multistage}
\vspace{-0.0mm}


%

We now propose a concept of a \emph{multi-stage} RE matching unit that uses aggressive approximate
reductions, which do not preserve the language of the NFA, to utilise FPGA resources
efficiently.
The architecture of the RE matching unit is composed of several stages (see
Fig.~\ref{fig:multistage} for an example of a~3-stage architecture).
Every stage in the architecture contains an instantiation of the RE matching
engine described in~\secref{sec:reg_engine}.

The idea is that each of the stages will use different NFAs---starting with a
bigger number of smaller and imprecise NFAs and proceeding to smaller numbers of
larger but more precise NFAs---to decrease, in a~resource-efficient way, the
number of packets entering the subsequent stage.
Consider an~NFA~$\A$ that recognizes the language~$\lang$ defined by the REs in
a~given \snort module.
The first stage of the architecture contains many copies of a~small NFA~$\A_1$,
which over-approximates~$\lang$, i.e., apart from all packets in~$\lang$, it
also matches some packets not in~$\lang$.
All matched packets are then sent to Stage~2, which contains less copies of
a~larger NFA~$\A_2$, which over-approximates~$\lang$, but more precisely than
$\A_1$ (which is the reason it is larger---less precise approximations
of~$\lang$ obtained by our reduction are usually smaller than more precise ones).

The number of copies of~$\A_2$ can be smaller due to the fact that the traffic
entering it is just a~fraction of the input traffic since Stage~1 has removed
a~significant number of packets from further processing.
Each subsequent stage contains an even smaller number of even more precise (and,
therefore, larger) NFAs.
The final stage contains either copies of~$\A$, in which case the output of the
RE matching unit is exactly the packets from~$\lang$, or as precise
over-approximation of~$\lang$ as possible given the available resources, in
which case the last remaining false positives need to be removed in software.

Our approximate reduction methods (described in detail
in~\secref{sec:approx_red}) output a~Pareto frontier of NFAs $\bbA = \{\A_1,
\ldots,$ $\A_k\}$ obtained from the input NFA~$\A$.  Each NFA~$\A_i$ comes with
two parameters: (1)~its size given as the number of \emph{look-up tables} (LUTs)
obtained from its HW synthesis, denoted by the function $\lut: \bbA \to \bbN$, and
(2)~the probability that it accepts an input packet, denoted by the function
$\accpt: \bbA \to \rangeof{0,1}$.
%
%
Note that lower is better for both parameters.

Given the~set~$\bbA$, we aim at obtaining a~configuration of the multi-stage
architecture that is as small and precise as possible.
This gives rise to the following two optimisation problems:
\begin{itemize}
  \item $OPT_{RSC}$:
    minimise the amount of resources (denoted by the variable~$\varof{RSC}$)
    used by the RE matching unit given a~maximum speed of traffic on its
    output~$X$ and
  \item $OPT_{out}$:
    minimise the speed of the traffic on the output of the last $n$-th stage of
    the RE matching unit (denoted by the variable~$\varof{out$_n$}$) given
    a~fixed~amount~of~resources~$Y$.
\end{itemize}
%
%

The optimization problems can be formalised (and then solved by a constraint
solver) using the following constraints.

The first constraint formalises that, in an $n$-stage
architecture, the output of each stage (denoted using
a~real-valued variable $\varof{out$_{i}$}$) is the fraction of the unit's input
traffic~$\varof{out$_0$}$ given by the acceptance probability of the used
NFA~$\A_j$ (the 0/1 variable~$\varof{$_i$uses$_j$}$ denotes that Stage~$i$ uses
the NFA~$\A_j$):

\vspace{-1em}
\begin{flalign*}
  \forall 1 \leq i \leq n:\,
  \varof{out$_i$} = \varof{out$_{0}$} \cdot \sum_{\A_j \in \bbA}
  \varof{$_i$uses$_j$} \cdot \accptof{\A_j}. && (1)
\end{flalign*}

The second family of constraints formalises that every stage uses precisely one
version of the NFA from~$\bbA$:
\begin{flalign*}
  &&
  \forall 1 \leq i \leq n: \,
    1 =
    \sum_{\A_j \in \bbA}
    \varof{$_i$uses$_j$}. && (2)
\end{flalign*}

Finally, the third constraint formalises the total resources~$\varof{RSC}$ used by
the architecture:
\begin{flalign*}
  \varof{RSC}
  =
  \sum_{1 \leq i \leq n}
  \left\lceil \frac{\varof{out$_{i-1}$}}{\mathit{TP}}
  \right\rceil \cdot
  \sum_{\A_j \in \bbA}
  \varof{$_i$uses$_j$} \cdot
  \lutof{\A_j}. && (3)
\end{flalign*}
We suppose that $\mathit{TP}$ is the \emph{throughput} of the NFAs (obtained as
the multiple of the NFA's input bit-width and the clock frequency).
The total resources are then computed in such a~way that the $i$-th stage
uses~$\lceil\varof{out$_{i-1}$} / \mathit{TP}\rceil$ NFAs $\A_j$ (since it needs to be
able to process all of the output traffic from the previous stage), each taking
$\lutof{\A_j}$ LUTs.

The considered optimisation problems can then be formalised as follows:
\vspace{-0.5em}
\begin{align*}
OPT_{RSC}: &\mbox{ Minimise }   \varof{RSC} \mbox{ subject to } \varof{out$_n$}
\leq X, \\
OPT_{out}:  &\mbox{ Minimise }   \varof{out$_n$} \mbox{ subject to } \varof{RSC}
\leq Y.
\end{align*}
Here, $X$ is the maximal permissible output traffic, and $Y$ is the maximal number of LUTs available in the architecture.

\smallskip

\emph{Example:}
Consider the scenario with 100\,Gbps input delivered over a~512-bit interface
on 200\,MHz.
Moreover, assume that the data width of the NFAs is set to 32~bits, their
throughput is, therefore, 6.4\,Gbps, so 16 of them are needed to process the
input 100\,Gbps, and, further, assume that there are 10,000 available LUTs.
Suppose that we are given a~precise NFA~$\A$ for
%
%
which our reduction
procedure---based on a sample of network traffic---yields a~set of approximated
NFAs $\bbA = \{\A_1, \A_2, \A_3 = \A\}$ (i.e., $\A_3$ is the precise NFA~$\A$)
such that their parameters are as given in Table~Ia.
Ignoring the (negligible) overhead of stage interconnection, the possible configurations of the multi-stage unit are presented in Table~Ib.
Note that Configuration 1 (which uses a~single stage with no
approximation) cannot fit within the available resources.
\qed
%
\begin{table}[t]
  \centering

  \caption{(a) Parameters of the NFAs. (b) Possible configurations of the multi-stage unit (output of each configuration is 10\,Gbps).}
  \label{tab:params-merge}
  \vspace{-4mm}
  \subfloat[\label{tab:nfa-params}]
  {
  \begin{tabular}{|m{3mm}||r|r|}
     \hline
     & $\lut$ & $\accpt$ \\
     \hline
     $\A_1$ & 100 & 0.5 \\
     $\A_2$ & 200 & 0.2 \\
     $\A_3$ & 1,000 & 0.1\\
     \hline
     \multicolumn{3}{r}{}
   \end{tabular}
  }
  \subfloat[\label{tab:stage-config}]
  {
    \begin{tabular}{|m{0.5mm}|c|c|c||r|}
      \hline
      \# & Stg.~1          & Stg.~2         & Stg.~3         & LUTs                  \\
      \hline
      1 & $16{\times}\A_3$ & ---             & ---             & \red{16,000} \\
      2 & $16{\times}\A_2$ & $4{\times}\A_3$ & ---             & 7,200        \\
      3 & $16{\times}\A_1$ & $8{\times}\A_3$ & ---             & 9,600        \\
      4 & $16{\times}\A_1$ & $8{\times}\A_2$ & $4{\times}\A_3$ & 7,200        \\
      \hline
    \end{tabular}
  }
\end{table}
%

\vspace{-0.0mm}
\section{Approximate NFA Reduction}\label{sec:approx_red}
\vspace{-0.0mm}
A~\emph{nondeterministic finite automaton} (NFA) over a finite alphabet $\Sigma$
is a~quadruple $\A = (Q,\trans,q_I,F)$ where $Q$~is a~finite set of
\emph{states}, $\trans \subseteq Q\times \Sigma \times Q$ is a~\emph{transition
relation}, $q_I \in Q$ is the \emph{initial state}, and $F\subseteq Q$ is a set
of $\emph{final states}$.
For  $q \in Q$ and $a \in \Sigma$, we write
$\delta(q,a)$ to denote the set
$\{ q' \in Q \mid (q,a,q') \in \delta \}$.
It is lifted to sets $S\subseteq Q$ as $\trans(S,a) = \bigcup_{q\in S}\trans(q,a)$
and to words as
$\extrans(q,\varepsilon) = \{q\}$ and $\extrans(q,wa) = \delta(\extrans(q,w),a)$ for $w\in\Sigma^*$.

Our hardware architecture uses NFAs to accept packets based on their prefixes,
and so we use a slightly non-standard notion of the \emph{language} of an NFA.
Namely, the language of a state $q$ is the set of words $\langA(q) = \{ w_1.w_2
\in \Sigma^* \mid \exists q_f \in F:~ q_f \in\extrans(q,w_1) \}$, i.e., words
whose prefix can take $\A$ from $q$ to some accepting state, followed by an
arbitrary suffix.
The language of $\A$ is then defined as $L(\A) = \langA(q_I)$.

\vspace{-0.0mm}
\subsection{Pruning Reduction}\label{sec:prun_red}
\vspace{-0.0mm}

Our first NFA reduction is the so-called \emph{pruning reduction}.
The reduction removes from the automaton a set $R$ of states considered as
\emph{insignificant}, together with all their adjacent transitions.
At the same time, in order to overapproximate the original language, all states
that are not removed and that have a transition going to a removed state are
made final.
Below, we call such states \emph{border states}, forming a set $B$.


More precisely, let $\A = (Q,\trans,q_I,F)$ be an NFA over~$\Sigma$,
and let  $R \subseteq Q$, $q_I \not\in R$, be a set of states to be removed (we
will later discuss how to find such a set).
Let $B = \{ q \in Q \setminus R \mid \exists a \in \Sigma:~ \trans(q,a) \cap R
\neq \emptyset \}$ be the set of border states corresponding to $R$.
The operation of pruning from $\A$ the states from $R$ produces the NFA
$\prunedA = (Q' = Q \setminus R,\trans', q_I, F')$ where $\delta' = \delta \cap (Q'
\times \Sigma \times Q')$ and $F' = (F \cap Q') \cup B$.

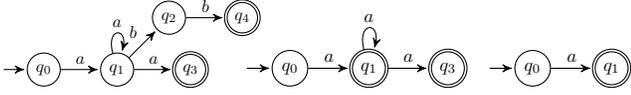
\begin{figure}[t]
\begin{tikzpicture}[->,>=stealth',shorten >=1pt,auto,node distance=1.5cm,
                    scale = 0.65,transform shape,initial text={}]
  \tikzstyle{every state}=[inner sep=3pt,minimum size=5pt]

  \node[state,initial] (q_0) {$q_0$};
  \node[state] (q_1) [right of=q_0] {$q_1$};
  \node[state] (q_2) [above right of=q_1] {$q_2$};
  \node[state,accepting] (q_3) [right of=q_1] {$q_3$};
  \node[state,accepting] (q_4) [right of=q_2] {$q_4$};

  \path (q_0) edge              node {$a$} (q_1)
        (q_1) edge[loop above]  node {$a$} (q_1)
        (q_1) edge[above left]       node {$b$} (q_2)
        (q_2) edge[above]       node {$b$} (q_4)
        (q_1) edge[above]       node {$a$} (q_3);

\end{tikzpicture}
\hspace*{-6mm}
\begin{tikzpicture}[->,>=stealth',shorten >=1pt,auto,node distance=1.5cm,
                    scale = 0.7,transform shape,initial text={}]
  \tikzstyle{every state}=[inner sep=3pt,minimum size=5pt]

  \node[state,initial] (q_0) {$q_0$};
  \node[state,accepting] (q_1) [right of=q_0] {$q_1$};
  \node[state,accepting] (q_3) [right of=q_1] {$q_3$};

  \path (q_0) edge              node {$a$} (q_1)
        (q_1) edge[loop above]  node {$a$} (q_1)
        (q_1) edge[above]       node {$a$} (q_3);

\end{tikzpicture}
\begin{tikzpicture}[->,>=stealth',shorten >=1pt,auto,node distance=1.5cm,
                    scale = 0.7,transform shape,initial text={}]
  \tikzstyle{every state}=[inner sep=3pt,minimum size=5pt]

  \node[state,initial] (q_0) {$q_0$};
  \node[state,accepting] (q_1) [right of=q_0] {$q_1$};

  \path (q_0) edge              node {$a$} (q_1);

\end{tikzpicture}
\caption{An example of pruning: the original NFA (left), after pruning (middle),
and after a subsequent simulation reduction (right).}
\label{fig:pruning}
\end{figure}

The pruning reduction over-approximates the
original language, i.e., $L(\A)\subseteq L(\prunedA)$.
The obtained NFA can, of course, be potentially further reduced by exact,
simulation-based reductions \cite{Reduce}.
An example of the pruning reduction is shown in Fig.~\ref{fig:pruning} assuming
that $R=\{q_2,q_4\}$ and hence $B=\{q_1\}$. Recall that the acceptance
is based on accepting prefixes.

The trade-off between reduction and accuracy that the pruning reduction offers
depends on the choice of the set $R$ of the states to be removed.
We therefore try to compose $R$ from such states of $\A$ that have the least
influence on the acceptance/rejection of packets in typical traffic.
For that, we use a representative sample $\calS$ of the network traffic in the
form of a multiset of packets.
We label each state $q \in Q$ of the input NFA $\A$ by its \emph{significance}:
the number  $\lab(q)$ of packets from $\calS$ over which the state $q$ can be
reached in $\A$ (if there are multiple ways of reaching $q$ over the same
packet, we do not distinguish them).
Formally, $\lab(q) = \sum_{w \in \{ w_1.w_2 \in \calS \mid q \in\extrans(q_I,w_1)
\}} \calS(w)$ where $\calS(w)$ is the number of occurrences of the packet $w$ in
the multiset $\calS$.

The error caused by the pruning reduction based on a~set of states $R$ wrt a
sample $\calS$ can be bounded in terms of significance of the border states $B$
corresponding to $R$.
Indeed, only the packets accepted at some border state can get wrongly accepted.
Formally, $\errorAc(\calS, \A, \prunedA) \leq \sum_{q \in B} \lab(q)$ where
$\errorAc(\calS, \A, \prunedA) = \sum_{w \in L(\prunedA)\setminus L(\A)}
\calS(w)$ is the exact error caused by the reduction on the sample $\calS$.

To specify the desired reduction, we use a target \emph{reduction ratio} $\theta
\in (0,1]$ meaning that the automaton should be reduced to $m = \lceil \theta
\cdot |Q| \rceil$ states.
To obtain $m$ states while minimising the error, we
fill $R$ with $|Q|-m$ least significant states\footnote{
This strategy can cause a~larger error when an originally
non-accepting border state of a high significance is forced to become
accepting by some insignificant accepting successor state.
This could be avoided, e.g.,
by preferring pruning
final states without a significant successor border state.
In our experiments, we, however, sufficed with the simple~strategy.
}.

Continuing with our example from Fig.~\ref{fig:pruning}, the reduction ratio is
$\theta = 0.6$.
To get the set $R = \{ q_2, q_4 \}$, the significance of the states $q_2$ and
$q_4$ must be smaller than that of $q_0$, $q_1$, and $q_3$.
Intuitively, this means that seeing $b$ after the initial $a$ symbols is rare.
Hence, even if we allow the automaton to accept a~string with the initial $a$
symbols followed by a~$b$~symbol that is not followed by the second required
$b$, no big error will be caused.
The fact that such scenarios are commmon for NFAs obtained from real-world REs
over real-world traffic is illustrated by our successful experiments
(\secref{sec:eval}).

\begin{figure}[t]
   \centering
   \includegraphics[keepaspectratio,width=88mm]{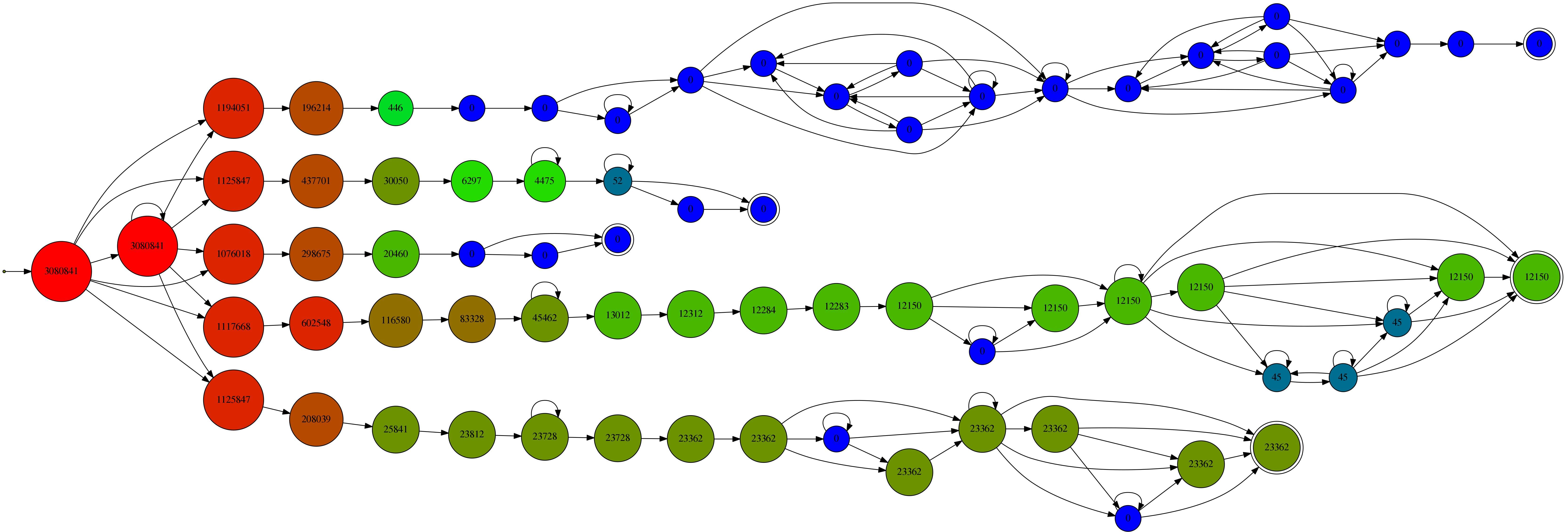}
   \vspace*{-6mm}
   \caption{A heat map illustrating significance of states in a typical NFA.}
   \vspace*{-4mm}
   \label{fig:heat}
\end{figure}

As an additional illustration, Fig.~\ref{fig:heat} shows a heat map of the
states of an NFA obtained from one simplified RE (the \texttt{sprobe} PCRE
mentioned in Sect.~\ref{sec:eval}), having a typical stucture of many of the
NFAs that one obtains from real-world PCREs, which was labelled over a sample of
real-world traffic.
The ``cold'' states (blue, green) have low significance and are good candidates
for pruning.

The significance of all states is computed efficiently in time $\bigO(kn^2)$
where $n=|Q|$ and $k =\sum_{w\in\calS}|w|\cdot\calS(w)$ is the overall length of
$\calS$.
For that, we can use the subset construction known from determinisation of NFAs,
just run on particular packets $w \in \calS$.
%
Namely, each state's significance is initially set to zero.
Then, we run~$\A$ over every $w = a_1 \ldots a_l \in \calS$, computing
consecutive sets of states $Q^i$ that are possibly reached after processing the
prefix $a_1 \ldots a_i$, and, at the end of the run, we increment by one the
significance of each of the states encountered on the way.
Formally, for each $w \in \calS$, we start with $Q^0  = \{q_I\}$, and,
subsequently, compute $Q^{i+1} = \delta(Q_1^i,a_{i+1})$ for all $1<i<l$.
The significance of all states in the set $\bigcup_{i=0}^lQ^i$ is then
incremented by one.

\vspace{-0.0mm}
\subsection{Merging Reduction}\label{sec:merge_red}
\vspace{-0.0mm}

Our second reduction, called a \emph{merging reduction}, is motivated by an
observation that, in typical traffic,
packets that start with a prefix of a~certain kind (i.e., they are from some
language $L$) almost always continue by an infix $w$ that follows
a~predetermined pattern (a~concrete word or a sequence of characters from
predetermined character classes).
We say that, in a sample $\calS$, the pattern of $w$ is \emph{predetermined} by
$L$.
The part of the automaton that, after reading the prefix from $L$, tests whether
the infix fits the pattern can be significantly simplified by collapsing it into
a single state with a self-loop over all the symbols that label the original
transitions while causing a small error only:
Indeed, it is unlikely that a~packet with an infix other than the predetermined
one will appear after the given prefix.
Note that the pruning reduction discussed previously is not suitable for
simplifying the states that test the pattern as they may be of an arbitrarily
high significance.

The operation of merging a sub-automaton based on a~set~$S$ of
states means to
(1) redirect the targets of transitions entering~$S$ to a~new state~$s$,
(2) reconnect all transitions leaving~$S$ to start from $s$ instead,
(3) make $s$ final iff any of the states in $S$ is final, and
(4) remove the states of $S$.
Note that, like pruning, merging also over-approximates the language
 by allowing any permutation of the infix pattern.

Our detection of the parts of automata to be merged---typically, sequences of
states---is based on a notion of \emph{distance} defined
for a~pair of states~$q$ and~$r$
as $\distance(q,r) =
\max\left(\frac{\lab(r)}{\lab(q)},\frac{\lab(q)}{\lab(r)}\right)$
if they are neighbours (i.e., $r\in\delta(q,a)$ or
$q\in\delta(r,a)$ for some $a$)
and as $\distance(q,r) = \infty$ otherwise.
%
Intuitively, a small $\distance(q,r)$ means that $\lab(q)$ and $\lab(r)$ are
similar, which typically happens if most of the packets reaching $q$ continue to
$r$ or vice versa.\footnote{In theory, it does not have to be the case, as
$\distance(q,r)$ may be polluted by packets reaching and leaving $q$ and $r$
from and to other states, but it is mostly the case in practice.}
Symbols on transitions from $q$ to $r$ hence form a predetermined pattern of
length 1.
Therefore, merging $q$ and $r$, and thus over-approximating the pattern, should
cause a small error only.
Merging of longer patterns is then achieved by merging multiple patterns of
length 1.

The merging reduction is parameterised by a \emph{distance ceiling}
$\mdistance$---we merge states with distance below $\mdistance$.
%
Formally, the sets of states to be merged are defined as the equivalence classes
of the smallest equivalence ${\disteq}\subseteq{ Q\times Q}$ that
contains all pairs $(q,r)$ with $\distance(q,r)\leq \mdistance$ (in other words,
$\disteq$ is the reflexive transitive closure of
$\{(q,r)\mid\distance(q,r)\leq \mdistance\}$).

The merging reduction does not provide theoretical guarantees, and it is to a
large degree based on empirical experience, in which its parametrisation with
$\mdistance$ allows one to control the ratio between reduction and error well.
There are, however, cases in which merging leads to an undesirable loss of
precision even with a~small $\mdistance$.
To limit such effects, we restrict merging by an additional parameter, the
\emph{frequency ceiling} $\mfrequency\in(0,1]$.
We prohibit merging of states with $\frequency(q) = \frac{\lab(q)}{|\calS|} >
\mfrequency$, that is, those whose frequency in $\calS$ is larger than the
ceiling.
Formally, given $\mdistance$ and $\mfrequency$, we merge the equivalence classes
of ${\disteq}\cap {\{(q,r)\mid \frequency(q)\leq \mfrequency \land
\frequency(r)\leq \mfrequency\}}$.

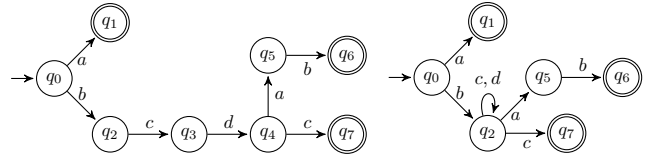
\begin{figure}[t]
\centering
\begin{tikzpicture}[->,>=stealth',shorten >=1pt,auto,node distance=1.5cm,
                    scale = 0.7,transform shape,initial text={}]
  \tikzstyle{every state}=[inner sep=3pt,minimum size=5pt]

  \node[state,initial] (q_0) {$q_0$};
  \node[state,accepting] (q_1) [above right of=q_0] {$q_1$};
  \node[state] (q_2) [below right of=q_0] {$q_2$};
  \node[state] (q_3) [right of=q_2] {$q_3$};
  \node[state] (q_4) [right of=q_3] {$q_4$};
  \node[state] (q_5) [above of=q_4] {$q_5$};
  \node[state,accepting] (q_6) [right of=q_5] {$q_6$};
  \node[state,accepting] (q_7) [right of=q_4] {$q_7$};

  \path (q_0) edge[below]       node {$a$} (q_1)
        (q_0) edge[above]       node {$b$} (q_2)
        (q_2) edge[above]       node {$c$} (q_3)
        (q_3) edge[above]       node {$d$} (q_4)
        (q_4) edge[right]       node {$a$} (q_5)
        (q_4) edge[above]       node {$c$} (q_7)
        (q_5) edge[below]       node {$b$} (q_6);

\end{tikzpicture}
\begin{tikzpicture}[->,>=stealth',shorten >=1pt,auto,node distance=1.5cm,
                    scale = 0.7,transform shape,initial text={}]
  \tikzstyle{every state}=[inner sep=3pt,minimum size=5pt]

  \node[state,initial] (q_0) {$q_0$};
  \node[state,accepting] (q_1) [above right of=q_0] {$q_1$};
  \node[state] (q_2) [below right of=q_0] {$q_2$};
  \node[state] (q_5) [above right of=q_2] {$q_5$};
  \node[state,accepting] (q_6) [right of=q_5] {$q_6$};
  \node[state,accepting] (q_7) [right of=q_2] {$q_7$};

  \path (q_0) edge[below]       node {$a$} (q_1)
        (q_0) edge[above]       node {$b$} (q_2)
        (q_2) edge[loop above]  node {$c,d$} (q_5)
        (q_2) edge[below]       node {$a$} (q_5)
        (q_5) edge[above]       node {$b$} (q_6)
        (q_2) edge[below]       node {$c$} (q_7);

\end{tikzpicture}
\vspace*{-2mm}
\caption{An input NFA (left) and an NFA obtained by merging (right).}
\label{fig:merging}
\end{figure}

An example of the merging reduction is shown in Fig.~\ref{fig:merging},
assuming  ${\distance(q_2,q_3) < D}$, $\distance(q_3,q_4) < D$, and that all
other distances are greater than~$D$.
To make
the states $q_2$, $q_3$, and $q_4$ less critical, we further assume
that $\frequency(q_2) < \mfrequency$, $\frequency(q_3) <
\mfrequency$, $\frequency(q_4) < \mfrequency$.
Intuitively, this means that only some packets continue over $b$ from $q_0$ to
$q_2$.
Roughly all of those packets then continue until they reach~$q_4$, where another
important split of the acceptance happens.
The prefix $b$, however, predetermines the subsequent occurrence of $cd$.
By simplifying the automaton and allowing any string from $\{c,d\}^*$ to appear
after $b$, no significant error arises since most packets will anyway contain
$cd$ after $b$ only.

%


\vspace{-0.0mm}
\section{Evaluation}\label{sec:eval}
\vspace{-0.0mm}

In this section, we present an experimental evaluation of the proposed approach on real RE-matching
instances.

\subsection{Considered REs}

%
\begin{table}[t]
  \caption{Sizes of the considered NFAs.}
  \label{tab:nfas-sizes}
  \centering
\vspace*{-2mm}
    \begin{tabular}{|l|r|r|}
      \hline
      NFA & States & Transitions\\
      \hline
      \texttt{backdoor} & 3,898 & 100,301 \\
      \texttt{l7-all} & 7,280 & 2,647,620 \\
      \texttt{pop3} & 923 & 209,467 \\
      \texttt{sprobe} & 168 & 5,108 \\
      \texttt{spyware} & 12,809 & 279,334 \\
      \hline
    \end{tabular}
\vspace*{-2mm}
\end{table}
We experimented with a~set of REs
describing protocols and attacks obtained
%
from the L7 classifier for the Linux
Netfilter\cite{netfilter} framework and from the \snort tool\cite{Snort}.
From the L7 classifier describing L7 protocols, we used all rules, giving us a
set of REs denoted as \texttt{l7-all} below.
From the \snort tool, we used the following set of REs: \texttt{backdoor},
\texttt{pop3}, and \texttt{spyware-put} (abbreviated as \texttt{spyware} below),
describing attacks on selected protocols.
We also used nine rules, denoted as \texttt{sprobe}, proposed for lawful
interception in cooperation with our national police.
We used the \netbench tool~\cite{netbench11}
for (i)~translating REs to NFAs and (ii)~the synthesis of reduced NFAs to VHDL.
%
The sizes of the NFAs obtained by translating the considered REs are shown in Table~\ref{tab:nfas-sizes}.

\subsection{Evaluation Data}

The sample of network traffic that we used for our experiments was obtained from
two measuring points of a nation-wide Internet provider
%
%
%
connected to a~100\,Gbps backbone link.
%
%
%
%
The training data used for labelling the automata contained $\sim$1M
packets sampled from the captured traffic during the time of 19.5\,min
containing 509M packets.
The testing data used for the subsequent evaluation consisted of
$\sim$21M
packets sampled from the captured traffic containing $\sim$210M packets.
The testing data was sampled over the time of 105 hours (over 4\,days) such that  every hour 1M packets were captured.

\subsection{Evaluation Environment}

We implemented the proposed  techniques  in
a Python prototype~\cite{ahofa}.
%
%
%
In the experiments, we ran merging with the frequency
ceiling $F=0.1$ and the distance ceiling $D=1.005$.
%
%
%
FPGA synthesis was done using Xilinx Vivado v.2018.1.

\subsection{Running Time}

Our reduction techniques are light-weight, and, in
contrast to existing approaches~\cite{CeskaHHLV18}, they are capable of reducing
large NFAs appearing in real-world RE matching problems for real network
scenarios.
(Recall that the largest automata we consider have over 12k states or over 2.5M
transitions.)
Using the training data containing 1M packets, we needed about 15 min to derive
the state labelling function $\ell$ for the largest considered NFAs.
The runtime of the other parts of the reduction process was then negligible.
Also note that, for a given NFA, the labelling can be performed only once for
various values of the reduction ratio $\theta$.
%

\subsection{Research Questions}

We are interested in the following two key research questions related
to the proposed approach:\begin{itemize}

  \item[\textbf{R1:}] Are our reduction techniques able to provide useful
  trade-offs between the reduction error and the reduction ratio?

  \item[\textbf{R2:}] Can the reduced NFAs be compiled into a multi-stage
  architecture with throughput of 100\,Gbps and beyond?

\end{itemize}


\medskip
\noindent
\textbf{R1}: \emph{Reduction Trade-offs}
\smallskip

In our experiments with the reduction techniques, we consider both the pruning
and merging reductions.
The merging reduction is, however, always combined with a~subsequent pruning
reduction as a~standalone use of merging turned out not to be effective.
Moreover, as a baseline, we also consider a so-called \emph{bfs-reduction}. It
does not use any training traffic and works by simply pruning away
states that are far from  the initial state.
%
All of the approximate reductions are followed by the exact simulation-based
reduction \cite{ClementeM13} whenever the tool \reduce~\cite{Reduce}
implementing this reduction is capable of handling the approximate
NFA (it fails on large NFAs).

We consider two metrics characterizing the reduced automata.
The first metric is the \emph{acceptance precision} $\ap=\frac{\atp}{\afp+\atp}$
where $\atp$ denotes acceptance true positives (the packet is accepted and
should have been accepted) and $\afp$ denotes acceptance false positives (the
packet is accepted and should \emph{not} have been accepted).
This metric expresses the ratio of correctly accepted packets to all accepted
packets from the testing traffic sample and hence characterizes the error caused
by the approximation.
Our second metric is the \emph{acceptance probability}
$\prob=\frac{\atp+\afp}{|\calS|}$ (where $|\calS|$ denotes the size of the
input network traffic sample) that captures what fraction of the input
traffic is accepted by the reduced NFA and passed to the next stage, i.e., how
much the NFA reduces the flow of packets to be further processed.
$\prob$ is an important metric for building efficient multi-stage architectures.

\begin{figure}
  \vspace*{-4.0mm}
  \hspace*{-7.5mm}
  \subfloat[\label{fig:l7-all-ap} $\mathit{AP}$ for \texttt{l7-all}.]
  {
    \includegraphics[width=50mm]{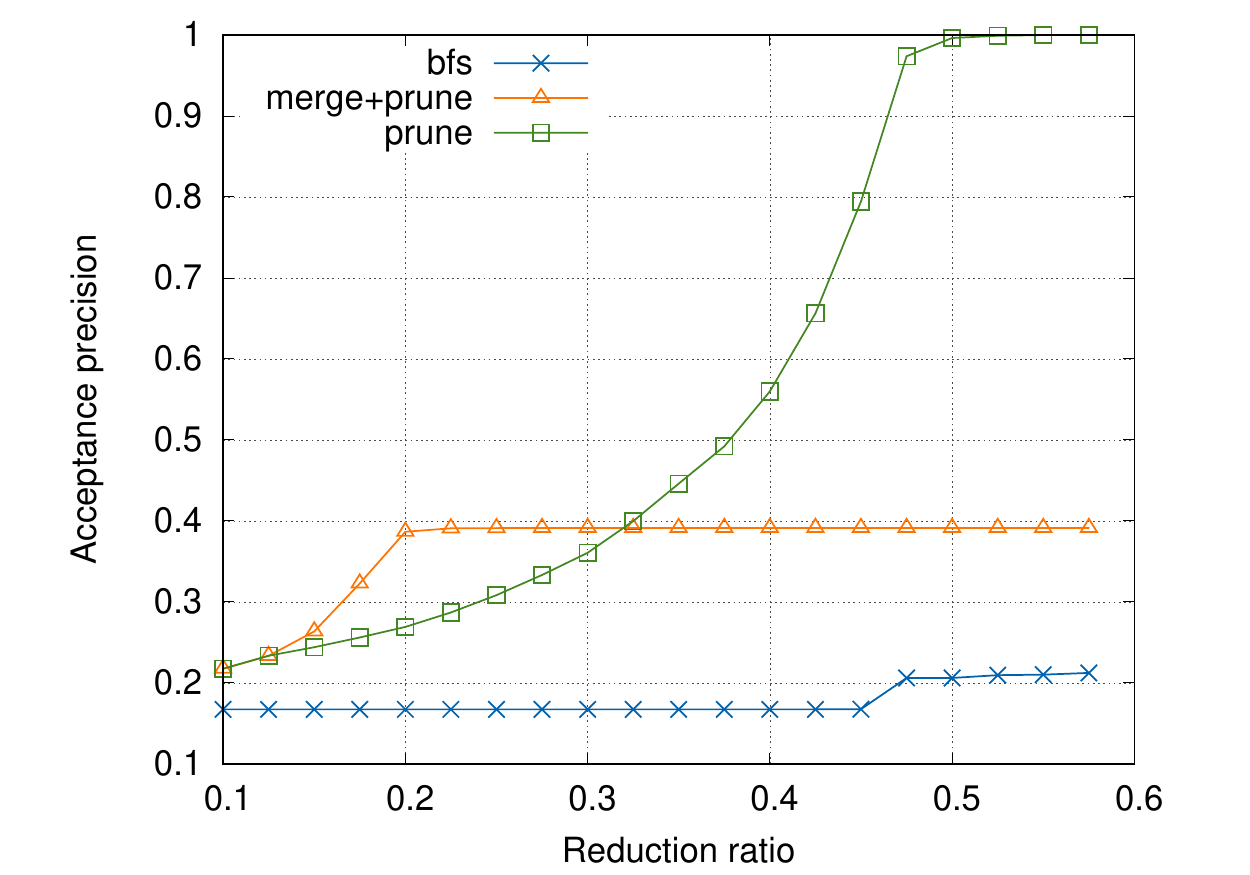}
  }
  \subfloat[\label{fig:l7-all-th} $\prob$ for \texttt{l7-all}.]
  {
    \hspace{-8mm}
    \includegraphics[width=50mm]{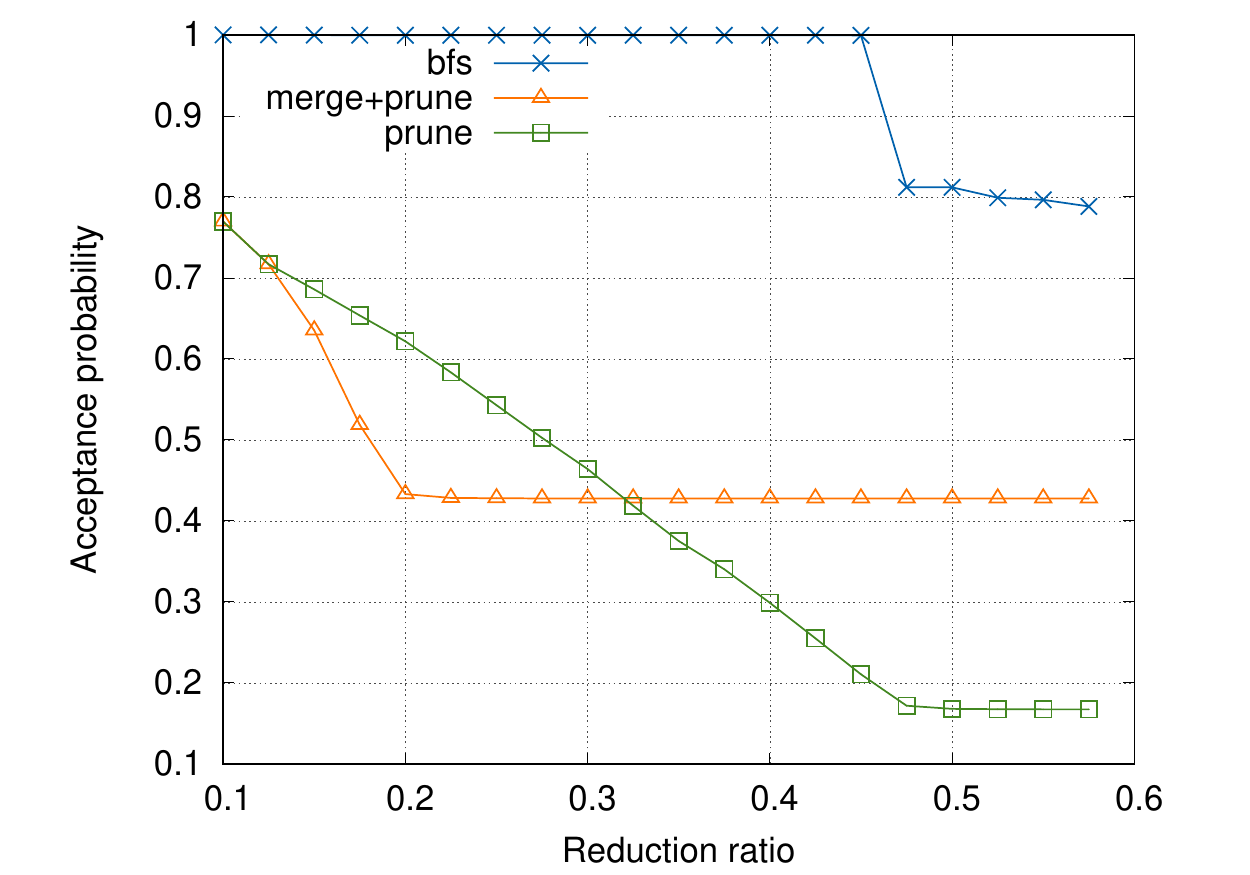}
  }
  \\[-2mm]
  \subfloat[\label{fig:backdoor-ap}$\mathit{AP}$ for \texttt{backdoor}.]
  {
    \hspace{-8mm}
    \includegraphics[width=50mm]{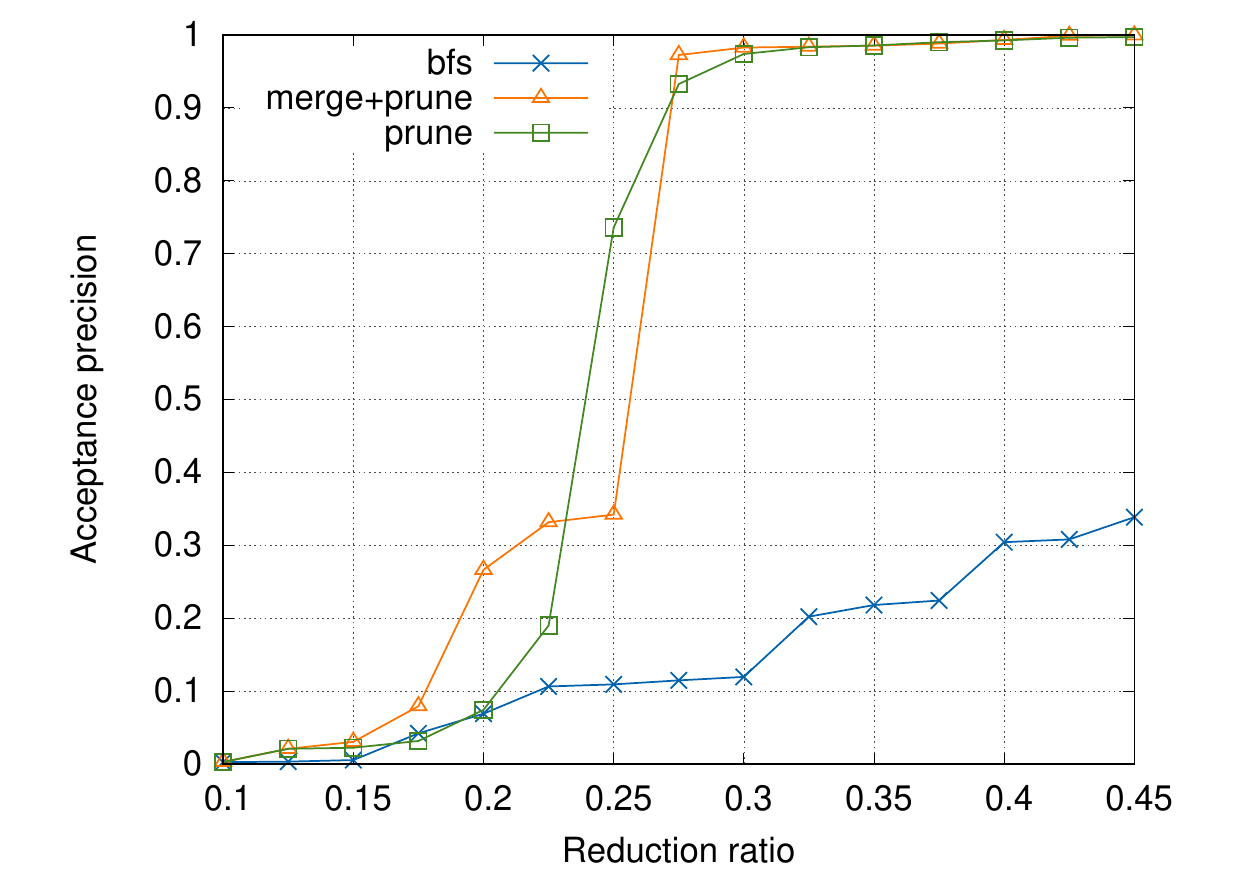}
  }
  \subfloat[\label{fig:backdoor-th}$\prob$ for \texttt{backdoor}.]
  {
    \hspace{-8mm}
    \includegraphics[width=50mm]{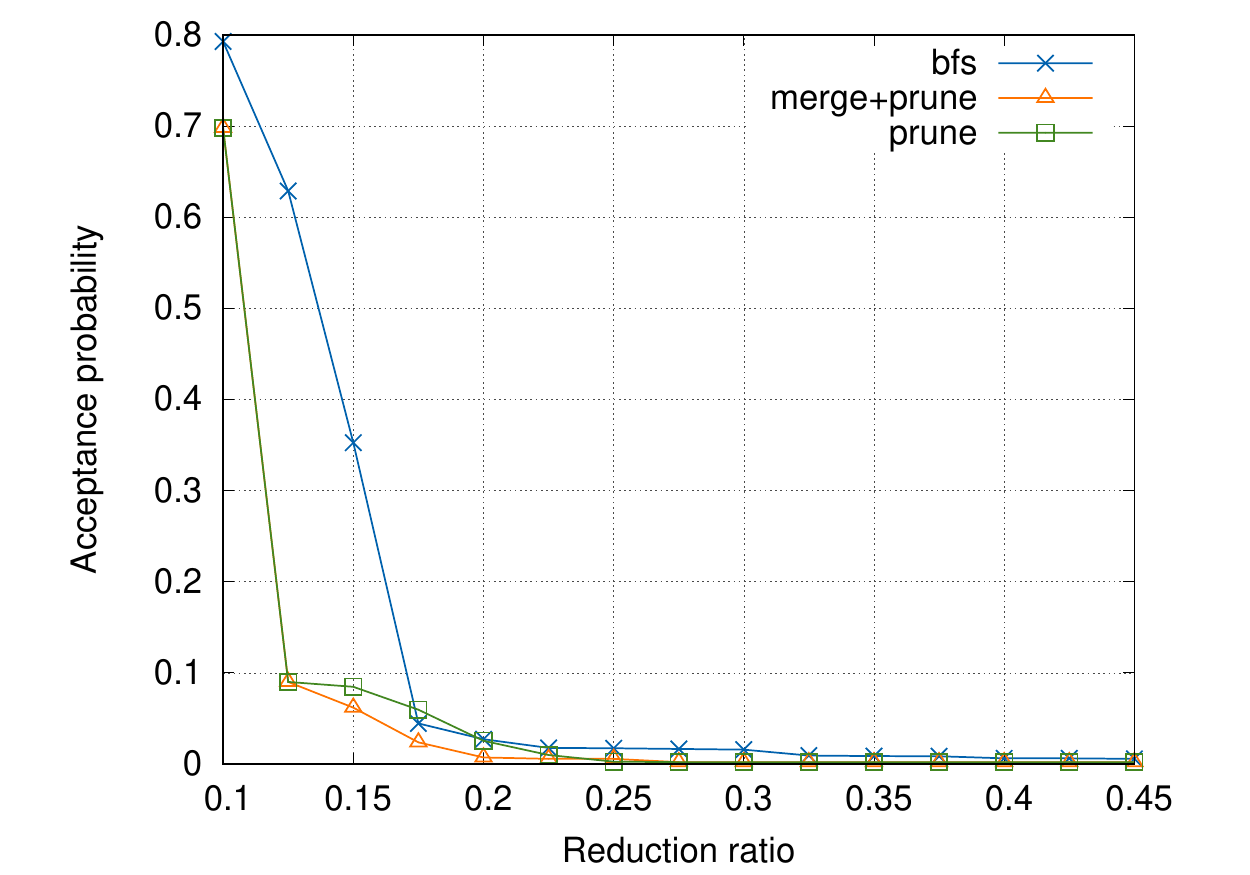}
  }
  \caption{Results for \texttt{l7-all} and \texttt{backdoor}.}
  \label{fig:l7-ap}
  \vspace{-3mm}
\end{figure}

\begin{figure}
  \centering
    \includegraphics[width=50mm]{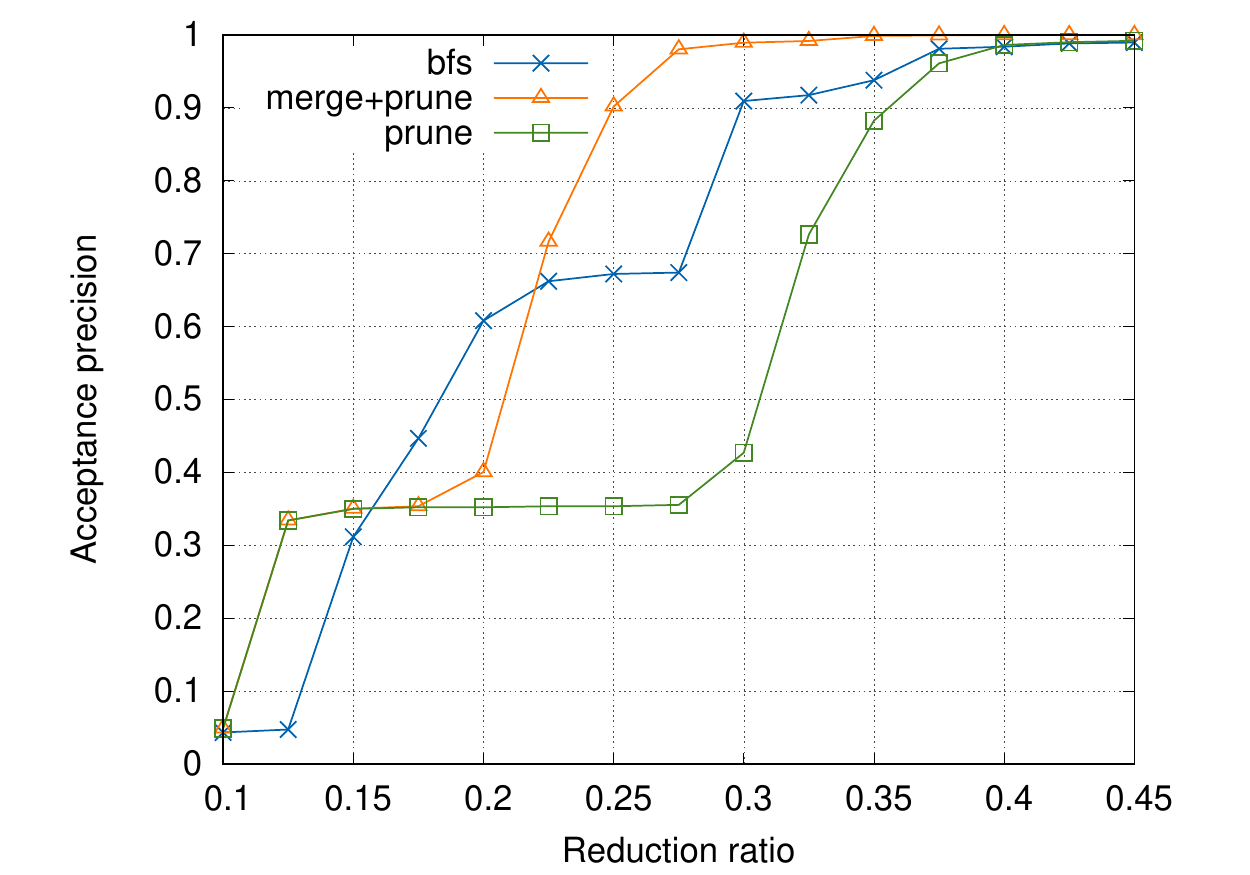}
  \vspace*{-3.0mm}
  \caption{$\mathit{AP}$ for \texttt{spyware}.}
  \label{fig:spyware}
  \vspace*{-2mm}
\end{figure}

Figs.~\ref{fig:l7-ap} and~\ref{fig:spyware} show the trade-offs achieved by our
different reduction strategies (namely, \emph{bfs}, \emph{prune}, and
\emph{merge-prune}) on challenging RE matching problems.

Figs.~\ref{fig:l7-all-ap} and~\ref{fig:l7-all-th}  show results for the NFA
describing \texttt{l7-all}.
We observe that the particular
reduction techniques provide a~different quality of the trade-offs.
In particular, \emph{bfs} is not capable of producing any useful approximation.
Further, we can observe that $\emph{merge-prune}$ dominates for reduction ratios
lower than 0.3, but it is significantly outperformed by $\emph{prune}$ for higher ratios.


The figures show that these trends are preserved for both 
metrics.
%
Note that the original NFA accepts around 17\,$\%$ of
the
traffic, and using the \emph{prune} technique, we obtain a~reduced NFA having
only a half of the states with almost the same acceptance probability~$\prob$.


%
The reduction trade-offs that we obtain for the NFA of the \texttt{backdoor}
attack are plotted in Figs.~\ref{fig:backdoor-ap} and~\ref{fig:backdoor-th}.
The \emph{bfs} reduction is again significantly outperformed by both the
$\emph{prune}$ and
%
%
$\emph{merge-prune}$ methods
when $\mathit{AP}$ is considered.
Note that these two techniques provide reductions that achieve almost a 100\,\%
$\mathit{AP}$ using only 35\,\% of the states of the original~NFA.
As the NFA accepts only 0.2\,\% of the traffic, we can obtain the accepting
probability $\prob$ that is close to this value using only 22\,\% of its
states regardless of the reduction used.
We observed similar trends also for the \texttt{pop3} and \texttt{sprobe}
attacks (not
presented here due to lack of space) where almost a 0\,\% $\prob$ was achieved using only
20\,\% and 25\,\% states, respectively.

Finally, we report the results for \texttt{spyware} REs in
Fig.~\ref{fig:spyware}.
We can see that, wrt $\mathit{AP}$, \emph{prune} lags behind the other two techniques for
reduction ratios between 0.15 and 0.35.
For higher reductions, all techniques provide $\mathit{AP}$ close to~100\,\%.
Similar trends can be observed also for the $\prob$ metric (not presented here). Note that the original NFA accepts about 3.5\,\% of the input traffic~only.


\vspace{0.5em}

The experiments conducted within \textbf{R1} clearly demonstrate that the
proposed reduction techniques are able to provide high-quality trade-offs
between the precision and the reduction factors. They also show that our
techniques can considerably outperform the baseline \emph{bfs-reduction} and
can handle very complex NFAs (having more than 10k states), where existing
methods (such as those in~\cite{CeskaHHLV18} and~\cite{Reduce}) fail.


\medskip
\noindent
\textbf{R2}: \emph{Compiling the Multi-Stage Architecture}
\smallskip

We will use the reduced NFAs from \textbf{R1} to obtain instantiations of the
multi-stage architecture that can be used in FPGA-based IDSes to effectively
decrease the amount of traffic that the software part of the IDS needs to
process.  We synthesise our designs for a~card with the Xilinx Virtex
UltraScale+ VU9P FPGA chip, which contains 1,182k\,LUTs (other resources are in
our case always dominated by the number of LUTs).  From our experience, it is
possible to use up to 70\,\% of the LUTs available on the FPGA and successfully
route designs at the considered frequency (200\,MHz), which leaves us with
827k\,LUTs that we can use.
%
%
Moreover, the components that we use for receiving packets and transfering them
to the CPU consume around 90k\,LUTs, we are therefore left with 737k\,LUTs for
RE matching.

In every stage of the RE matching unit, we use the pipelined NFAs
architecture described in~\secref{sec:reg_engine} instantiated with 8-bit data-width of the NFAs,
which gave us the best results.
Therefore, the throughput (parameter $\mathit{TP}$
in~\secref{sec:multistage}) of every NFA at 200\,MHz is 1.6\,Gbps.
The optimisation problem of compiling the multi-stage architecture leads to a mixed
integer quadratically constrained quadratic program, which we solve using the
Gurobi solver~\cite{gurobi} (this step took at most 2\,s).

Our goal is to obtain single-box IDSes using a~combination of hardware
preprocessing and a~software IDS.
This means that the task of the hardware accelerator is to decrease the amount
of the traffic entering the software part as much as possible while keeping all
suspicious packets.
Ideally, the final stage of the RE matching unit would be the precise NFA,
so the hardware accelerator would output precisely the packets that match
the given set of REs.
Due to the size of the precise NFAs, this is, however, often infeasible, so we
also consider the setting that decreases the traffic as much as possible.

Below, we provide results of our experiments for some of the considered sets of
REs.
We tried to compile architectures for the speeds 100, 200, and 400\,Gbps using
1--4 stages.

\vspace{-0.0mm}
\subsubsection{\texttt{backdoor}}\label{sec:label}
\vspace{-0.0mm}

In Table~\ref{tab:backdoor-stg}, we present results of optimal architectures for the
\snort's \texttt{backdoor} module. We
\begin{table}[t]
  \centering
  \caption{Optimal multistage architecture for the \texttt{backdoor} module.}
  \label{tab:backdoor-stg}
  \vspace*{-2mm}
 \setlength{\tabcolsep}{4.5pt}
   \begin{tabular}{|c||r|r|r|r|}
    \hline
    \multicolumn{5}{|c|}{Precise} \\
    \hline
     speed & 1\,stg & 2\,stg & 3\,stg & 4\,stg \\ \hline
     100   & 236k  & 56k   & 50k   & 50k  \\
     200   & 473k  & 113k  & 99k   & 96k   \\
     400   & \red{946k}  & 223k  & 194k  & 186k \\
     \hline
   \end{tabular}
\end{table}
present only results for the precise setting, as our multi-stage architectures
can handle 400\,Gbps traffic using the available resources (737k\,LUTs).
Note that the single stage architecture (i.e., the ``1\,stg'' column), can process
traffic up to 200\,Gbps only (the precise NFA consumes 3,695\,LUTs).
In order to process 400\,Gbps, it is necessary to use the multi-stage
architecture, in which case four stages give the best results.

\vspace{-0.0mm}
\subsubsection{\texttt{spyware}}\label{sec:label}
\vspace{-0.0mm}

Our results for the \snort's \texttt{spyware}
module are shown in Table~\ref{tab:spyware-stg}.
This module is much
\begin{table}[t]
 \vspace*{-2mm}
 \centering
 \caption{Optimal multistage architecture for the \texttt{spyware} module.}
 \label{tab:spyware-stg}
 \vspace*{-2mm}
 \setlength{\tabcolsep}{4.5pt}
\renewcommand{\arraystretch}{.96}
   \begin{tabular}{|c||r|r|r|r||r|r|r|r|}
    \hline
    \multicolumn{5}{|c||}{Precise} & \multicolumn{4}{c|}{4\,\% of traffic} \\ \hline
     speed & 1\,stg & 2\,stg & 3\,stg & 4\,stg & 1\,stg & 2\,stg & 3\,stg & 4\,stg \\ \hline
     100 & \red{5M}  & 444k & 296k & 296k &  227k & 61k  & 65k  & 69k  \\
     200 & \red{10M} & \red{809k} &  513k &  513k &  453k & 122k & 126k & 133k \\
     400 & \red{20M} & \red{1.5M}   & \red{945k} & \red{945k} &  \red{907k} & 242k & 247k & 261k
     \\
     \hline
   \end{tabular}
   \vspace*{-3mm}
\end{table}
more complex than \texttt{backdoor}, since its
precise NFA~takes $\sim$78k\,LUTs.
Therefore, for the throughputs of\linebreak 100\,Gbps and 200\,Gbps,
the multi-stage ar\-chi\-tec\-ture is needed.
For 400\,Gbps, we were not
able to obtain a~precise configuration; we were, however, able to obtain
multi-stage configurations (the best one with 2 stages) that decrease the
amount of traffic sent to the CPU below 4\,\% (i.e., 16\,Gbps). Although
16\,Gbps is on the edge of\linebreak capabilities of current
SW-based IDSes, we stress that the\linebreak final-stage reduced NFA
has $\mathit{AP}$ very close to 100\,\% (Fig.~\ref{fig:spyware}) and thus
only a~small fraction of packets are misclassified.
\vspace{-0.0mm}
\subsubsection{\texttt{l7-all}}\label{sec:label}
\vspace{-0.0mm}


Our most challenging example is from the \texttt{l7-all} RE set.
Although the size of the precise automaton is
not as large~as~for \texttt{spyware} (the precise NFA
consumes 27,650 LUTs), it is less amenable for approximate reduction
\begin{table}[t]
  \centering
  \caption{Optimal multistage architecture for the \texttt{l7-all} module.}
  \label{tab:l7-all-stg}
  \vspace*{-2mm}
   \setlength{\tabcolsep}{4.5pt}
\renewcommand{\arraystretch}{.96}
    \begin{tabular}{|c||r|r|r|r||r|r|r|r|}
      \hline
      \multicolumn{5}{|c||}{Precise} & \multicolumn{4}{c|}{17\,\% of traffic} \\
      \hline
       speed & 1\,stg & 2\,stg & 3\,stg & 4\,stg & 1\,stg & 2\,stg & 3\,stg & 4\,stg \\ \hline
       100 & \red{1.8M}  & \red{894k} & \red{880k} & \red{880k} &  \red{1.1M} & 597k & 648k  & 701k \\
       \hline
     \end{tabular}
   \vspace*{-3mm}
\end{table}
%
because, in contrast to \snort modules, it contains REs that
are matched by many
packets. The results for the \texttt{l7-all} RE set are shown in Table~\ref{tab:l7-all-stg}. Our best solution reduces the
input traffic from 100\,Gbps to 17\,Gbps
and uses 597\,kLUTs in
two stages. As in the previous case,
the final-stage reduced NFA has almost 100\,\%  precision and thus
only a~small fraction of packets are misclassified.


\vspace{-0.0mm}
\subsubsection{\texttt{sprobe} and \texttt{pop3}}\label{sec:label}
\vspace{-0.0mm}

The sets of REs for \texttt{sprobe} and \texttt{pop3}  are, on the other hand,
quite less challenging.  The precise NFAs consume only 195 and
1,721\,LUTs, respectively, so we can easily obtain a~precise design at 400\,Gbps
with only a~single stage using $\sim$50k and $\sim$440k\,LUTs,
respectively.


\vspace{0.5em}

The experiments conducted within \textbf{R2} clearly demonstrate the practical
potential of our approach.
The key observation is that the resource reductions provided by the particular
multi-stage architectures directly depend on the characteristics of the
underlying NFAs (both the precise NFA and the reduced variants) and the typical
traffic.
Apart from the size of the precise NFA, there are two crucial characteristics:
(1)~whether the number of packets accepted by the precise NFA is low and (2)
whether  the reduction can compress the NFA while not increasing the number of
accepted packets too much.
%
%
If both these conditions are met (as for \texttt{backdoor} and
\texttt{spyware}), we observe drastic resource savings allowing us to achieve
throughput of the resulting IDSes going beyond 100\,Gbps,
which is unprecedented for REs of such size and complexity.
On the other hand, if the original NFA is large, accepts many packets, and
highly precise reductions achieve only moderate reductions (as for
\texttt{l7-all}), the resulting multi-stage architecture provides only moderate
savings and ensuring 100\,Gbps remains at the edge of what we can achieve.

\vspace{-0.0mm}
\section{Conclusion}\label{sec:concl}
\vspace{-0.0mm}

    We have leveraged techniques for approximate reduction of
    NFAs to allow RE matching of a~set of \snort/Netfilter modules on speeds significantly beyond the capabilities of the state-of-the-art
    single-box solutions, namely 100 and even 400\,Gbps.
    The use of the approximate reduction allowed us to significantly decrease the
    size of the NFAs while keeping the number of false positives low (e.g.
    for \snort's \texttt{spyware} module, we obtained a~reduction to 28\,\% of
    the original size while keeping the error below 2\,\%).
    Moreover, the use of the multi-stage architecture
    better utilises FPGA resources (e.g., using 3 stages, we were able to reduce
    the resources consumed by \texttt{spyware} down to $\sim$5\,\% of the
    original without introducing any error).
    As far as we know, our technique is the first solution that can be used to
    obtain single-box IDSes detecting large sets of complex REs at speeds over
    100\,Gbps.

\subsubsection*{Acknowledgement}

We thank Vlastimil Ko\v{s}a\v{r} for his comments on an earlier draft of the
paper and Martin \v{Z}\'{a}dn\'{i}k for providing us with the backbone network
traffic.
This work was supported by The Ministry of Education, Youth and Sports from the
National Programme of Sustainability (NPU~II) project IT4Innovations
excellence in science --- LQ1602.

%
%




\balance

\end{document}